# Common Signal Analysis

Chengpu Wang        Independent researcher        631-974-1078        Chengpu@gmail.com

## 1   Abstract

A common signal is defined for any two signals which have non-zero correlation. A mathematical method is provided to extract the best obtainable common signal between the two signals. This analysis is extended to extracting common signal among three signals.

## 2   Introduction

When two signals have non-zero correlation [1], it is commonly believed that these two signals have similarity between them. However, so far there is no clear definition about what the similarity actually means. This paper defines the similarity as common signal between the two signals, and provides a mathematical way to extract it.

## 3   Common Signal Analysis between Two Signals

### 3.1   Definition of Common Signal

Let $\alpha$ be a constant. Let $\sigma^2()$, $\mathrm{Cov}()$ and $\gamma()$ be variance, covariance and correlation functions. Assume A, $B_1$, and $B_2$ are three mutually independent random variables. Define

$$\beta_1{}^2 \equiv \frac{\sigma^2(B_1)}{\sigma^2(A)} \text{ and } \beta_2{}^2 \equiv \frac{\sigma^2(B_2)}{\sigma^2(\alpha A)} = \frac{\sigma^2(B_2)}{\alpha^2 \sigma^2(A)}.$$ Define $\sigma^2 \equiv \sigma^2(A)$. Define \$x as sign of value x.

Equation 3-3 [2] gives the correlation $\gamma_{12}$ between two signals $S_1 \equiv A + B_1$ and $S_2 \equiv \alpha A + B_2$.

Equation 3-1:        $\sigma_1{}^2 \equiv \sigma^2(S_1) = \sigma^2(A + B_1) = (1 + \beta_1{}^2)\sigma^2$ ;

Equation 3-2:        $\sigma_2{}^2 \equiv \sigma^2(S_2) = \sigma^2(\alpha A + B_2) = (1 + \beta_2{}^2)\alpha^2\sigma^2$ ;

Equation 3-3: $\gamma_{12} = \dfrac{Cov(S_1, S_2)}{\sigma_1 \sigma_2} = \dfrac{\alpha \sigma^2}{\sigma_1 \sigma_2} = \dfrac{\$\alpha}{\sqrt{1 + \beta_1{}^2}\sqrt{1 + \beta_2{}^2}}$ ;

Equation 3-3 can also be interpreted reversely: if two signals $S_1$ and $S_2$ are correlated by a non-zero $\gamma_{12}$, each signal can be viewed as containing a completely uncorrelated part and a completely correlated part. The completely correlated part is defined as the common signal between the two signals $S_1$ and $S_2$. In this context, $\alpha$ measures relative strength of the common signal between $S_1$ and $S_2$, while $\beta_1$ and $\beta_2$ measure the background strength within $S_1$ and $S_2$ respectively. The question is how to extract the common signal.

## 3.2  Best Obtainable Common Signal

The correlation between A and $S_1$ or $S_2$ are $\gamma_1$ or $\gamma_2$ respectively.

Equation 3-4:     $\gamma_1 \equiv \dfrac{Cov(S_1, A)}{\sigma_1 \sigma} = \dfrac{\sigma}{\sigma_1} = \sqrt{\dfrac{1}{1 + \beta_1{}^2}}$ ;

Equation 3-5:     $\gamma_2 \equiv \dfrac{Cov(S_2, A)}{\sigma_2 \sigma} = \dfrac{\alpha \sigma}{\sigma_2} = \$\alpha \sqrt{\dfrac{1}{1 + \beta_2{}^2}} = \dfrac{\gamma_{12}}{\gamma_1}$

Compose a new signal $S_- \equiv \alpha S_1 - S_2 = \alpha B_1 - B_2$, which contains no common signal A. Compose another new signal $S_a \equiv S_1 - \chi_1 S_-$ and choose $\chi_1$ to maximize the correlation $\gamma_a$ between A and $S_a$. The optimal result $\chi_1$, $\gamma_a$ and $S_a$ are shown by Equation 3-7.

Equation 3-6:     $1/\gamma_a{}^2 \equiv \dfrac{\sigma^2(S_a)\sigma^2(A)}{Cov(S_a, A)^2} = 1 + (1 - \alpha \chi_1)^2 \beta_1{}^2 + \alpha^2 \chi_1{}^2 \beta_2{}^2$ ;

Equation 3-7:     $\chi_1 = \dfrac{\beta_1{}^2}{\beta_1{}^2 + \beta_2{}^2} \dfrac{1}{\alpha}$ :

$\gamma_a{}^2 = \gamma_{best}{}^2 \equiv \dfrac{\beta_1{}^2 + \beta_2{}^2}{\beta_1{}^2 + \beta_2{}^2 + \beta_1{}^2 \beta_2{}^2} \geq \gamma_1{}^2, \gamma_2{}^2$     $S_a = S_{best} \equiv \dfrac{\beta_2{}^2 S_1 + \beta_1{}^2 S_2 / \alpha}{\beta_1{}^2 + \beta_2{}^2}$ ;

$S_{best}$ can also be shown to have minimal correlation with $S_-$.

Similarly, compose a new signal $S_b \equiv S_2 + \chi_2 S_-$ and choose $\chi_2$ to maximize the correlation $\gamma_b$ between X and $S_b$. The optimal result $\chi_1$ and $S_a$ are shown by shown by Equation 3-9.

Equation 3-8: $\qquad 1/{\gamma_b}^2 \equiv \dfrac{\sigma^2(S_b)\sigma^2(X)}{Cov(S_b, X)^2} = 1 + {\chi_2}^2 {\beta_1}^2 + (1-\chi_2)^2 {\beta_2}^2 \ ;$

Equation 3-9: $\qquad \chi_2 = \dfrac{{\beta_2}^2}{{\beta_1}^2 + {\beta_2}^2} : \qquad \gamma_b = \gamma_{best} \qquad S_b = \alpha S_{best} \ ;$

Because the optimal $S_a$ and $S_b$ are completed correlated, they are the best obtainable common signals between the two correlated signals $S_1$ and $S_2$ in terms of a linear combination of $S_1$ and $S_2$.

## 3.3   Common Signal Extraction

In addition to $S_1$ and $S_2$, only $\gamma_{12}$, $\sigma_1$ and $\sigma_2$ are measurable. $\gamma_1 = \sigma/\sigma_1$ is used as an indeterminate variable whose range is $[|\gamma_{12}|, 1]$. $\alpha$, $\beta_1$ and $\beta_2$ can be expressed using $\gamma_1$, $\gamma_{12}$, $\sigma_1$ and $\sigma_2$ so that Equation 3-7 is translated into Equation 3-10 and Equation 3-11. To accommodate when $S_1$ and $S_2$ may not be directly comparable, Equation 3-11 is in the format of analyzing two unit-less signal $S_1/\sigma_1$ and $S_2/\sigma_2$, with result $S_{best}$ in the unit of $S_1$.

Equation 3-10: $\qquad {\gamma_{best}}^2 = \dfrac{{\gamma_{12}}^2 - 2{\gamma_1}^2{\gamma_{12}}^2 + {\gamma_1}^4}{{\gamma_1}^2(1-{\gamma_{12}}^2)} \ ;$

Equation 3-11: $\qquad \dfrac{S_{best}}{\sigma_1} = \dfrac{\dfrac{S_1}{\sigma_1}{\gamma_1}^2({\gamma_1}^2 - {\gamma_{12}}^2) + \dfrac{S_2}{\sigma_2}\gamma_{12}{\gamma_1}^2(1-{\gamma_1}^2)}{{\gamma_1}^2(1-{\gamma_{12}}^2) + {\gamma_{12}}^2(1-{\gamma_1}^2)} \ ;$

Additional condition is required to find the best obtainable common signal, such as a prior knowledge of either $\gamma_1$ or $\gamma_2$, or $\sigma$ in unit of either $S_1$ or $S_2$, or $\alpha$, or the ratio between ${\beta_1}^2$ and ${\beta_2}^2$. For example, when $\sigma = \sigma_1$, $\gamma_1 = 1$, $\gamma_2 = \gamma_{12}$, $\gamma_{best} = 1$ and $S_{best} = S_1$. If there is prior knowledge of characteristics of the common signal instead, a series of $S_{best}$ can be generated for different $\gamma_1$, and the selected $S_{best}$ determines $\gamma_1$. If there is no ground to judge which of $S_1$ and $S_2$ contains more common signal, assume $\gamma_1 = \gamma_2$, and the solution is Equation 3-12. Equation 3-12 is also the solution for minimizing $\gamma_{best}^2$, when the uncertainty of $\gamma_1$ has least effect on the result.

Equation 3-12: $\quad |\gamma_1| = |\gamma_2| = \sqrt{|\gamma_{12}|} \;\leq\; \gamma_{best} = \sqrt{\dfrac{2|\gamma_{12}|}{|\gamma_{12}|+1}} \;:\; \dfrac{S_{best}}{\sigma_1} = \dfrac{1}{2}\left(\dfrac{S_1}{\sigma_1} + \$\gamma_{12}\dfrac{S_2}{\sigma_2}\right);$

In Equation 3-12 $S_{best}$ is a surprisingly simple weighted average of $S_1$ and $S_2$. It can be explained well in two extreme cases when $S_1$ and $S_2$ are directly comparable:

- When signal strengths are same, $S_{best}$ gives heavier weight for signal which has less background strength.
- When background strengths are same, $S_{best}$ scales $S_1$ and $S_2$ according to their signal strengths.

## 3.4  Validation

A sine waveform is chosen to be the common signal A. A cosine waveform of twice the periodicity of A is chosen to be $B_1$, which is the background for signal $S_1$. A white noise is chosen to be $B_2$, which is the background for signal $S_2$. The white noise $B_2$ tends to have small but non-zero correlation with either the common signal A or the background $B_1$, and it is one of the major source when the result is deviate from ideal behavior. Equation 3-12 is used to extract $\gamma_{best}$ and $S_{best}$ regardless what $\alpha$, $\beta_1$ and $\beta_2$ are used to generate the signals. Table 1 shows that the correlation $\gamma_{best}$ to the true common signal A of $S_{best}$ in most case is better than the corresponding values $\gamma_1$ and $\gamma_2$. Figure 1 shows one of the exceptional case in which $\gamma_{best}$ is worse than $\gamma_1$, when $\alpha = 2$, $\beta_1 = 0.5$, $\beta_2 = 2$. It shows that contrary to smaller $\gamma_{best}$, $S_{best}$ is visually close to A than $S_1$, e.g., no more double positive peaks, and smaller negative peaks. It is observed that $S_{best}$ generally has better or at least comparable visual correlation to the true common signal A than $S_1$ or $S_2$.

## 4 Common Signal Analysis among Three Signals

### 4.1 Ideal Extraction

For each signal, assume j=1,2,3, $B_j$ is the background, $\sigma_j$ is the deviation, $\alpha_j$ is the signal strength, and $\beta_j$ is the background strength. Let $\alpha_1 = 1$. Assume A is the common signal, so that by definition $Cov(A, B_j) = 0$

Equation 4-1: $\qquad \sigma_j{}^2 \equiv \sigma^2(S_j) = \sigma^2(\alpha_j A + B_j) = (1 + \beta_j{}^2)\alpha_j{}^2 \sigma^2 \ ;$

Equation 4-2: $\qquad \gamma_j \equiv \dfrac{Cov(S_j, A)}{\sigma_j \sigma} = \dfrac{\alpha_j \sigma}{\sigma_j} = \$\alpha_j \sqrt{\dfrac{1}{1 + \beta_j{}^2}} \ ;$

Three signals can be composed according to Equation 4-3, each of which contains no common signal. However, only two of them are independent. So the signal $S_a$ to be optimized for highest correlation with common signal A is given by Equation 4-4 and simplified as Equation 4-5.

Equation 4-3: $\qquad$ N$_{12}$ ≡ α$_2$S$_1$−α$_1$S$_2$; $\qquad$ N$_{13}$ ≡ α$_3$S$_1$−α$_1$S$_3$; $\qquad$ N$_{23}$ ≡ α$_3$S$_2$−α$_2$S$_3$;

Equation 4-4: $\qquad$ S$_a$ ≡ S$_1$ − χ$_2$N$_{12}$ −χ$_3$N$_{13}$ = (α$_1$−χ$_2$α$_2$−χ$_3$α$_3$)S$_1$ + χ$_2$S$_2$ + χ$_3$S$_3$;

Equation 4-5: $\qquad$ S$_a$ = A + (α$_1$−χ$_2$α$_2$−χ$_3$α$_3$)B$_1$ + χ$_2$B$_2$ + χ$_3$B$_3$;

Using assumption of Equation 4-6, Equation 4-9 and Equation 4-10 show that both signal strength set {α$_2$, α$_3$} and background strength set {β$_1$$^2$, β$_2$$^2$, β$_3$$^2$} are completely solved by the correlation set {γ$_{12}$, γ$_{13}$, γ$_{23}$} (which is given by Equation 4-7) when Equation 4-8 is satisfied.

Equation 4-6: i ≠ j: $\qquad$ Cov(B$_i$, B$_j$) = 0;

Equation 4-7: i ≠ j: $\qquad \gamma_{ij} = \dfrac{Cov(S_i, S_j)}{\sigma_i \sigma_j} = \dfrac{\$\alpha_i \$\alpha_j}{\sqrt{1 + \beta_i{}^2}\,\sqrt{1 + \beta_j{}^2}} \ ;$

Equation 4-8: $\qquad \gamma_1{}^2 = \left| \dfrac{\gamma_{12}\gamma_{13}}{\gamma_{23}} \right| \le 1; \qquad \gamma_2{}^2 = \left| \dfrac{\gamma_{12}\gamma_{23}}{\gamma_{13}} \right| \le 1; \qquad \gamma_3{}^2 = \left| \dfrac{\gamma_{13}\gamma_{23}}{\gamma_{12}} \right| \le 1;$

Equation 4-9:    $\beta_j{}^2 = \dfrac{1}{\gamma_j{}^2} - 1\,;$

Equation 4-10:    $\alpha_2 = \$\gamma_{12}\left|\dfrac{\gamma_2\sigma_2}{\gamma_1\sigma_1}\right|\,;$      $\alpha_3 = \$\gamma_{13}\left|\dfrac{\gamma_3\sigma_3}{\gamma_1\sigma_1}\right|\,;$

The optimal signal and its correlation to the common signal A are provided by Equation 4-11 and Equation 4-12.

Equation 4-11:    $S_{best} \equiv \dfrac{\beta_2{}^2\beta_3{}^2 S_1/\alpha_1 + \beta_1{}^2\beta_3{}^2 S_2/\alpha_2 + \beta_1{}^2\beta_2{}^2 S_3/\alpha_3}{\beta_1{}^2\beta_2{}^2 + \beta_1{}^2\beta_2{}^2 + \beta_2{}^2\beta_3{}^2}\,;$

Equation 4-12:    $\gamma_{best}{}^2 \equiv 1 - \dfrac{\beta_1{}^2\beta_2{}^2\beta_3{}^2}{\beta_1{}^2\beta_2{}^2 + \beta_1{}^2\beta_2{}^2 + \beta_2{}^2\beta_3{}^2 + \beta_1{}^2\beta_2{}^2\beta_3{}^2}\ \ge\ \gamma_1{}^2, \gamma_2{}^2, \gamma_3{}^2\,;$

Like in the two-signal case, other optimized constructions of linear combination of $S_1$, $S_2$ and $S_3$ also result in Equation 4-11 and Equation 4-12, which are symmetric for each signal in the set.

When Equation 4-8 is satisfied, all correlations can be attributed to the common signal among three signals, and there is a unique and determinate solution for the best obtainable common signal $S_{best}$ among signal set $\{S_1, S_2, S_3\}$.

## 4.2  Non-Ideal Extraction

When Equation 4-8 cannot be satisfied, Equation 4-6 can not be satisfied either, and each $\gamma_{i,j}$ corresponds to an indeterminate variable $\mathrm{Cov}(B_i, B_j)$.  There are three more measurements $\{\sigma_1, \sigma_2, \sigma_3\}$, but five more indeterminate variables $\{\alpha_2, \alpha_3\}$ and $\{\beta_1{}^2, \beta_2{}^2, \beta_3{}^2\}$.  In addition, $\{B_1, B_2, B_3\}$ set as a whole shall not contribute to the extracted common signal.  How to translate the last requirement into mathematical formula is not clear at the moment.  It is even possible that no common signal exists for the set, and a new standard is required to judge for this condition.

For a set of N signals, if Equation 4-6 still holds, then there will be N(N-1)/2 count of Equation 4-7 but only N count of $\beta_j$, so that $\{\beta_j\}$ set is over-determinate.  Thus Equation 4-6 can not be satisfied generally when N > 3, and non-ideal extraction is needed for obtaining common signal from the signal set.

## 5   Summary and Discussion

The common signal is defined between two signals which have non-zero correlation. It is also defined among three signals whose correlations satisfy Equation 4-8. When it is defined, it can be approached by the best obtainable common signal, which is an optimal linear combination of the original signal set. The best obtainable common signal is under-determinate for the two signal set, and ideally determinate for the three signal set.

The common signal has not been well defined in other cases. If common signal is an intrinsic property of the signal set in these cases, the best obtainable common signal should not depends on the sequence of extracting it. Perhaps a common signal can be defined and extracted using this principle in non-ideal cases.

Applying ideal extraction repeatedly may obtain the common signal among a set of $3^n$ signals Ideal conditions. Another approach when N > 3 is to see if the N(N-1)/2 count of Equation 4-7 degenerate into N independent equations using solutions similar to Equation 4-8, Equation 4-9 and Equation 4-10 which are obtained from a subset of Equation 4-7. If so, ideal extraction may be still possible.

## 7  Tables and Figures

| $\alpha$ | $\beta_1$ | $\beta_2$ | $\gamma_1$ | $\gamma_2$ | $\gamma_{best}$ |
|---|---|---|---|---|---|
| 2 | 1 | 1 | 0.7071 | 0.7652 | 0.8119 |
| 2 | 0.5 | 0.5 | 0.8944 | 0.9218 | 0.9393 |
| 2 | 2 | 2 | 0.4472 | 0.5106 | 0.5773 |
| 2 | 0.5 | 2 | 0.8944 | 0.5106 | 0.8936 |
| 2 | 2 | 0.5 | 0.4472 | 0.9218 | 0.8131 |
| 1 | 2 | 0.5 | 0.4472 | 0.9218 | 0.8852 |

Table 1: $\gamma_{best}$ vs. $\gamma_1$ and $\gamma_2$ for different $\alpha$, $\beta_1$ and $\beta_2$.

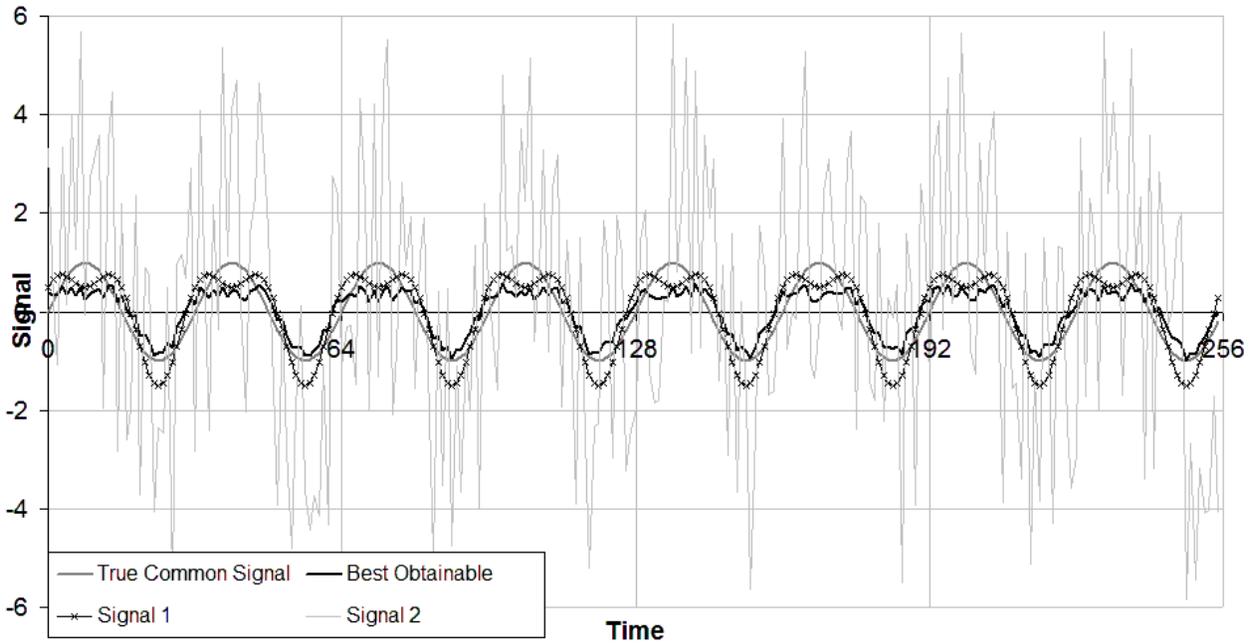

Figure 1: The true and best obtainable common signals vs. the original signals.